\documentclass[10pt]{iopart}
\usepackage[sort]{cite}
\usepackage{graphicx}
\usepackage{braket}

\begin{document}
\title[Rydberg excitons in electric and magnetic fields]%
{Rydberg excitons in electric and magnetic fields obtained with
  the complex-coordinate-rotation method}

\author{Patrik Zielinski, Patric Rommel, Frank Schweiner and\\
  J\"org Main}

\address{Institut f\"ur Theoretische Physik 1, Universit\"at Stuttgart,
  70550 Stuttgart, Germany}
\ead{main@itp1.uni-stuttgart.de}

\begin{abstract}
The complete theoretical description of experimentally observed
magnetoexcitons in cuprous oxide has been achieved by F.~Schweiner
\etal [Phys.\ Rev.\ B \textbf{95}, 035202 (2017)], using a complete
basis set and taking into account the valence band structure and the
cubic symmetry of the solid.  Here, we extend these calculations by
investigating numerically the autoionising resonances of cuprous oxide
in electric fields and in parallel electric and magnetic fields
oriented in [001] direction.  To this aim we apply the
complex-coordinate-rotation method.  Complex resonance energies are
computed by solving a non-Hermitian generalised eigenvalue problem,
and absorption spectra are simulated by using relative oscillator
strengths.  The method allows us to investigate the influence of
different electric and magnetic field strengths on the position, the
lifetime, and the shape of resonances.
\end{abstract}
\vspace{2pc}
\vfill
\submitto{\jpb}
\maketitle
\ioptwocol

\section{Introduction}
\label{sec:introduction}
Excitons are quasi particles, which occur in semiconductors and
insulators.
If an electron is raised from the valence band to the conduction band,
the remaining positively charged hole in the valence band interacts
with the negatively charged electron in the conduction band.
This electron-hole pair is called an exciton.
Depending on the spatial distance between electron and hole one
distinguishes between Frenkel and Mott-Wannier excitons
\cite{Mott38,Wan37}.
Frenkel excitons are confined to one lattice atom, whereas
Mott-Wannier excitons extend over many unit cells and can be treated
approximately as a hydrogenlike system.

An ideally suitable crystal for the experimental investigation of
Rydberg excitons is cuprous oxide (Cu$_2$O), where excitons
have been observed up to principal quantum number $n=25$
\cite{Kaz14,Thewes15}.
This has opened the field of research of giant Rydberg excitons.
As a consequence of the non-parabolic valence band structure of
Cu$_2$O, the simple hydrogenlike model does not describe the exciton
spectra very well \cite{Schoene16,Schweiner16b}.
This is especially true for excitons in external electric or magnetic
fields.
Heck\"otter \etal\cite{Heck17a} have investigated the influence of
different (weak) electric fields on the transmission spectra and have
shown that the transmission spectra depend on the crystal orientation
and the light polarisation.
Schweiner \etal\cite{Schweiner17a} have calculated the absorption
spectra of magnetoexcitons for various magnetic field strengths by
using a complete basis set and considering the complex valence band
structure.
The detailed comparison between the experimental and theoretical
spectra shows excellent agreement.
Similar is true for exciton spectra in the Voigt configuration, where
the external magnetic field is perpendicular to the incident light and
a weak effective electric field perpendicular to the magnetic field is
induced by the Magneto-Stark effect \cite{Rommel18}.

The experiments and calculations mentioned above are restricted to
bound states, and the experimentally observed linewidths are dominated
by exciton-phonon interactions \cite{Schweiner16a,Stolz18}.
However, by applying an external electric field, the potential barrier
of the Coulomb potential is lowered.
The electron can tunnel, and former bound states become quasi-bound or
resonance states.
They can be described by complex energies, where the imaginary part is
related to the decay rate and thus the linewidth of the resonance state.

The dissociation of excitons in Cu$_2$O by an electric field has been
investigated by Heck\"otter \etal\cite{Heck18}.
It has been shown that, similar to the Stark effect in atoms, the
field strength for dissociation decreases with increasing principal
quantum number $n$, but increases, for fixed $n$, with growing exciton
energy.
The experimental results have been compared with a theoretical
computation based on a simplified hydrogen-like model neglecting spin,
spin-orbit, and exchange interactions.

In the present paper we want to go beyond these calculations and
investigate the unbound resonance states of excitons in electric
fields or combined electric and magnetic fields by fully including the
effects of the valence band.
To this aim we extend the method introduced in
\cite{Schweiner16b,Schweiner17a} for the computation of exciton
spectra using a complete basis set, by the method of complex
coordinate-rotation \cite{Rei82,Ho83,Moi98}, which transforms the
Hermitian Hamiltonian with real eigenvalues to a non-Hermitian
operator with possibly complex eigenvalues.
The complex coordinate-rotation is a well established technique for
the computation of resonances in atomic physics, and has already been
applied, e.g., to the hydrogen atom in external fields
\cite{Mai92,Mai94,Car07,Car09}.
Here, we will calculate the positions of excitonic resonances in the
complex plane.
In particular, we will discuss the appearance and position of resonance
states depending on the electric field strength in Faraday
configuration, where the external field is parallel to the incident light.
Additionally, we will investigate the behaviour of resonance states in
parallel electric and magnetic fields.
We are also able to calculate
directly the relative oscillator strength, e.g., for $\sigma^+$ and
$\sigma^-$ polarised light and to simulate the corresponding
absorption spectra.

The paper is organised as follows:
In section~\ref{sec:theory} we present the theory.
After the introduction of resonance states and the complex
coordinate-rotation-method in section~\ref{sec:resonance} we present
in section~\ref{sec:hamiltonian} the Hamiltonian for the yellow
excitons in Cu$_2$O taking into account the non-parabolic valence band
structure and the effects of the external electric and magnetic fields.
In section~\ref{sec:eigenvalues} we discuss the setup of the
non-Hermitian generalised eigenvalue problem by using a complete basis set.
Formulas for the calculation of the relative oscillator strength and
the simulation of the absorption spectra are derived in
section~\ref{sec:Oscillatorstrength}.
The results of our calculations are presented in
section~\ref{sec:results}, and conclusions are drawn in
section~\ref{sec:conclusion}.

\section{Theory}
\label{sec:theory}
For the convenience of the reader we briefly recapitulate the
complex-coordinate-rotation method and the Hamiltonian of cuprous
oxide in external fields.
We then discuss the setup of a non-Hermitian generalised eigenvalue
problem for the computation of the complex resonance energies and the
corresponding eigenstates, and finally present the necessary equations
for the calculation of the oscillator strengths and the simulation of
the absorption spectra.

\subsection{Complex coordinate-rotation}
\label{sec:resonance}
Resonances are quasi-bound states with a finite lifetime.
Simple examples are the radioactive decay of unstable atomic nuclei,
an excited atom returning to its ground state, or a temporally
trapped particle in an open potential.
In the case of excitons the potential barrier of the Coulomb
interaction between electron and hole can be lowered by an external
electric field.
To describe the temporal decay of such systems we use non-Hermitian
Hamiltonians obtained with the method of the complex
coordinate-rotation \cite{Rei82,Ho83,Moi98}.

To introduce the formalism we calculate the energy expectation value
of the 1S state of the hydrogen atom,
\begin{equation}
  \braket{E}=\frac{\int_{0}^{\infty} R(r)\left[-\frac{1}{2}\frac{1}{r^2}
      \frac{\rm{d}}{\rm{d}r}r^2\frac{\rm{d}}{\rm{d}r}-\frac{1}{r}\right]
    R(r)r^2\rm{d}r}{\int_{0}^{\infty} R^2(r)r^2 \rm{d}r} = -\frac{1}{2}
\label{eq:reeles_Linienintegral}
\end{equation}    
with radial function $R(r)=2e^{-r}$.
Since $R(r)$ is an analytic function we can use Cauchy's integral theorem
and rewrite the real line integral \eref{eq:reeles_Linienintegral} into
a complex one \cite{Rei82},  
\begin{equation}
  \braket{E}=\frac{\int_{C} R(z)\left[-\frac{1}{2}\frac{1}{z^2}
      \frac{\rm{d}}{\rm{d}z}z^2\frac{\rm{d}}{\rm{d}z}-\frac{1}{z}\right]
    R(z)z^2\rm{d}z}{\int_{C} R^2(z)z^2 \rm{d}z}=-\frac{1}{2}
\label{eq:komplex_Linienintegral}
\end{equation}
with $z=re^{i\theta}$ and $\theta$ the angle between real and
imaginary part.
As long as we integrate from $z=0$ to $\infty$ the integral
\eref{eq:komplex_Linienintegral} does not depend on the value of
$\theta$, because the wave function $R(r)$ vanishes sufficiently quickly for
$r\rightarrow\infty$.
Thus, the results of both integrals \eref{eq:reeles_Linienintegral}
and \eref{eq:komplex_Linienintegral} are the same, i.e.,
an important property of the method is that bound states are invariant
under complex rotation.

To illustrate what happens to the scattering and continuum states,
we substitute $r\to re^{i\theta}$ and $k\to ke^{-i\theta}$ in the
result of the radial scattering problem \cite{Moi98},
\begin{equation}
  \Psi_{\mathrm{scatt}}(r) = A(k)\frac{e^{ikr}}{r}+B(k)\frac{e^{-ikr}}{r} \, .
\label{eq:streuung}
\end{equation}
The complex rotation implies a change of the energy (with $m=1$, $\hbar=1$),
\begin{equation}
  E = \frac{1}{2} k^2 \rightarrow E = \frac{1}{2} k^2e^{-2i\theta} \, ,
\end{equation} 
which means that these states are now rotated into the lower half of
the complex plane by the angle $2\theta$.

The most important feature of the complex coordinate-rotation concerns
the resonance states.
For appropriately chosen $\theta$, they appear as new eigenvalues on
the lower half of the complex energy plane.
A simple example is the inverted harmonic oscillator, which has no
bound states.
The complex rotated Hamiltonian is given by \cite{Rei82}
\begin{equation}
  H_{\mathrm{rot}} = -\frac{1}{2}e^{-2i\theta}\frac{\rm{d}^2}{\rm{d}x^2}
  - \frac{1}{2}e^{2i\theta}x^2
\end{equation} 
with purely imaginary energy eigenvalues
\begin{equation}
E_n = -i\left(n+\frac{1}{2}\right) \, .
\end{equation}
The inverted harmonic oscillator thus has an infinite number of
resonances with different widths $\Gamma_n=-2\,\mathrm{Im}\,E_n=2n+1$.

\begin{figure}
\includegraphics[width=0.49\textwidth]{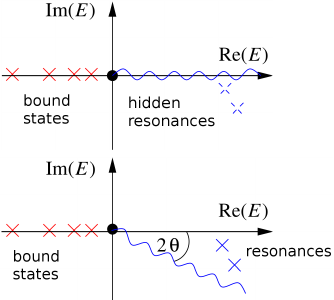}
\caption{Effect of the complex coordinate-rotation.  The hidden
  resonances are exposed by the rotation of the continuum states.}
\label{fig1:rotation}
\end{figure}
The results of the complex rotation can be summarised by the following
statements, which are also illustrated in figure~\ref{fig1:rotation}:
%\begin{itemize}
%\item
(i) The real-valued bound states are invariant under the complex rotation.
%\item
(ii) The energy values of the scattering respectively continuum states
are rotated into the lower half of the complex plane by $2\theta$.
% \item
(iii) For appropriately chosen angles $\theta$ resonances are exposed
by the rotation of the continuum states.
%\end{itemize}
    
\subsection{Hamiltonian of yellow excitons in external fields}
\label{sec:hamiltonian}
For the calculation of the yellow exciton series in Cu$_2$O we use the
same Hamiltonian as Schweiner \etal\cite{Schweiner16b,Schweiner17a,Schweiner17b}.
Without external fields the Hamiltonian can be written as
\begin{equation}
  H = E_{\rm g}+H_e(\bi{p}_{\rm e})+H_{\rm h}(\bi{p}_{\rm h})
  +V(\bi{r}_{\rm h}-\bi{r}_{\rm e}) \, ,
\label{eq:hamiltonian_without_fields}
\end{equation}
where the kinetic energy of the electron and hole are given by
\begin{eqnarray}
  H_e(\bi{p}_{\rm e}) &= \frac{\bi{p}_{\rm e}^2}{2m_{\rm e}} \, , \\
  H_{\rm h}(\bi{p}_{\rm h}) &= H_{\rm SO}+\frac{1}{2\hbar^2m_0}
  \{\hbar^2(\gamma_1+4\gamma_2)\bi{p}^2_{\rm h}\nonumber\\
&+2(\eta_1+2\eta_2)\bi{p}^2_{\rm h}(\bi{I}\cdot\bi{S}_{\rm h})\nonumber\\
&-6\gamma_2(p^2_{\rm h1}\bi{I}^2_1+{\rm c.p.})\nonumber\\
&-12\eta_2(p^2_{\rm h1}\bi{I}_1\bi{S}_{\rm h1}+{\rm c.p.})\nonumber\\
&-12\gamma_3(\{p_{\rm h1},p_{\rm h2}\}\{\bi{I}_1,\bi{I}_2\}+{\rm c.p.})\nonumber\\
&-12\eta_3(\{p_{\rm h1},p_{\rm h2}\}(\bi{I}_1\bi{S}_{\rm h2}
         +\bi{I}_2\bi{S}_{\rm h1})+{\rm c.p.})\} \, ,
\end{eqnarray}
with the spin-orbit interaction
\begin{equation}
H_{\rm SO}=\frac{2}{3}\Delta\left(1+\frac{1}{\hbar^2}\bi{I}\cdot\bi{S}_{\rm h}\right).
\end{equation}
Here, $E_{\mathrm{g}}$ is the gap energy,
$m_{\rm e}$ is the effective electron mass,
$\bi{p}=(p_1,p_2,p_3)$ are the momenta,
$\{a,b\}=\frac{1}{2}(ab+ba)$ is the symmetric product,
$\Delta$ is the spin-orbit coupling constant,
$\bi{I}$ is the quasispin,
$\bi{S}_{\rm h}$ is the hole spin, and
c.p.\ denotes cyclic permutation.
The vectors $\bi{I}$ and $\bi{S}_{\rm h}$ contain the components of
the three spin matrices $\bi{S}_{\rm{h}j}$ and $\bi{I}_j$ of the hole
spin $S_{\rm h}=1/2$ and the quasispin $I=1$, respectively.
The quasispin $\bi{I}$ is introduced to describe the degeneracy of the
valence band Bloch functions \cite{Luttinger56}.
The corresponding matrices fulfil the commutation relations of a spin
$I = 1$.
The parameters $\eta_j$ and the three Luttinger parameters
$\gamma_j$ are used to describe the behaviour and the anisotropic
effective hole mass in the vicinity of the $\Gamma$ point
\cite{Schweiner16b}.
The electric interaction between hole and electron is given by the
Coulomb potential
\begin{equation}
  V(\bi{r}_{\rm h}-\bi{r}_{\rm e}) =
  -\frac{e^2}{4\pi\epsilon_0\epsilon|\bi{r}_{\rm h}-\bi{r}_{\rm e}|}
\end{equation}    
with the dielectric constant $\epsilon$.
In references \cite{Schweiner17b,Schweiner17c} additional central cell
corrections were included in the Hamiltonian.
Since these effects are only important for states with principal
quantum numbers $n \leq 2$ \cite{Schweiner17b}, which we do not
consider in this paper, these corrections can be neglected here.

To take external electric and magnetic fields into account the
Hamiltonian~\eref{eq:hamiltonian_without_fields} must be extended.
The electric field is included by adding the potential
\begin{equation}
  V_{\rm F}(\bi{r}_{\rm h}-\bi{r}_{\rm e}) =
  -e(\bi{r}_{\rm h}-\bi{r}_{\rm e})\cdot \bi{F} \, ,
\end{equation}  
where $\bi{F}$ is the electric field vector.
To describe a constant magnetic field we use the vector potential with
symmetric gauge, $\bi{A}=(\bi{B}\times\bi{r})/2$.
The energy of the spins in the magnetic field is given by
\begin{equation}
  H_{\rm B} = \mu_{\rm B}\left[g_{\rm c}\bi{S}_{\rm e}
    + (3\kappa+g_{\rm s}/2)\bi{I}
    - g_{\rm s}\bi{S}_{\rm h}\right]\cdot\bi{B}/\hbar \, ,
\end{equation}
with $\mu_{\rm B}$ the Bohr magneton,
$g_{\rm s}$ the g factor of the hole spin $\bi{S}_{\rm h}$,
$g_{\rm c}$ the g factor of the electron spin $\bi{S}_{\rm e}$, and
$\kappa$ the fourth Luttinger parameter, which has been determined by
Schweiner \etal\cite{Schweiner17a}.
Next we introduce relative and centre of mass coordinates \cite{Schmelcher92},
\begin{eqnarray}
  \bi{r}&= \bi{r}_{\rm h}-\bi{r}_{\rm e}\, ,\quad
  \bi{R}=\frac{m_{\rm h}\bi{r}_{\rm h}+m_{\rm e}\bi{r}_{\rm e}}{m_{\rm h}+m_{\rm e}}\, ,\nonumber\\
  \bi{P}&= \bi{p}_{\rm h}+\bi{p}_{\rm e}\, ,\quad
  \bi{p}=\frac{m_{\rm h}\bi{p}_{\rm e}-m_{\rm e}\bi{p}_{\rm h}}{m_{\rm h}+m_{\rm e}} \, ,
\end{eqnarray}
and set the position and momentum of the centre of mass to zero
($\bi{R}=0$, $\bi{P}=0$).
The complete Hamiltonian of excitons in external fields with relative
coordinates finally reads
\begin{eqnarray}
  H &= E_{\rm g}+H_e(\bi{p}+e\bi{A}(\bi{r}))+H_{\rm h}(-\bi{p}+e\bi{A}(\bi{r}))\nonumber\\
    &+ V(\bi{r})+H_{\rm B}+V_{\rm F}(\bi{r}) \, .
\label{eq:Hamiltonianmitfelder}
\end{eqnarray}
More details of the derivations are given in references
\cite{Schweiner16b,Schweiner17a,Schweiner17c}.
The material parameters for Cu$_2$O used in our calculations are
listed in table~\ref{table1}.
\begin{table}
\renewcommand{\arraystretch}{1.2}
     \centering
     \caption{Material parameters of Cu$_2$O used in the calculations.}
     \begin{tabular}{l|l c}
     \hline
       Energy gap  & $E_{\rm g}=2.17208\,$eV & \cite{Heck17a}\\
       Effective electron mass  & $m_{\rm e}=0.99m_0$ & \cite{Hodby_1976} \\
       Effective hole mass  & $m_{\rm h}=0.58m_0$ & \cite{Hodby_1976} \\
       Dielectric constant      & $\epsilon=7.5$ &\cite{landolt1987landolt}\\
       Spin-orbit coupling       & $\Delta=0.131\,$eV &\cite{Schoene16}\\
       Valence band parameter & $\gamma_1=1.76$&\cite{Schoene16}\\
       ~~& $\gamma_2=0.7532$&\cite{Schoene16}\\
       ~~& $\gamma_3=-0.3668$&\cite{Schoene16}\\
       ~~&$\kappa=-0.5$& \cite{Schweiner17a}\\
        ~~& $\eta_1=-0.020$&\cite{Schoene16}\\
        ~~& $\eta_2=-0.0037$&\cite{Schoene16}\\
        ~~& $\eta_3=-0.0337$&\cite{Schoene16}\\
      g factor of the electron spin&$g_{\rm c}=2.1$&\cite{Schweiner17a}\\
      g factor of the hole spin&$g_{\rm s}\approx2$&\cite{Schweiner17a}
     \end{tabular}
\label{table1}
\end{table}

\subsection{Non-Hermitian generalised eigenvalue problem}
\label{sec:eigenvalues}
For the computation of eigenvalues of the yellow excitons in external
fields we now express the Hamiltonian~\eref{eq:Hamiltonianmitfelder} 
as a matrix by using an appropriate basis set.
For the radial part of the wave functions we use the Coulomb-Sturmian
functions \cite{Sturmbasis}
\begin{equation}
  U_{NL}(\rho) = N_{NL}(2\rho)^Le^{-\rho}L_N^{2L+1}(2\rho) \, ,
\label{eq:basis_r}
\end{equation}
where $L_N^{2L+1}$ are the associated Laguerre polynomials,
the $N_{NL}$ are normalisation factors, and
$\rho=r/\alpha$, with $\alpha$ being a free parameter.
$N$ is the radial quantum number, which is related to the principal
quantum number $n$ via $n = N + L + 1$.
Note that the Coulomb-Sturmian functions~\eref{eq:basis_r} form a
complete basis, however, they are not orthogonal.
For the computation of resonances the complex coordinate-rotation
$r\to r\e^{\rmi\theta}$ discussed in section~\ref{sec:resonance} is
equivalent to choosing the free parameter $\alpha$ as being complex,
i.e., $\alpha=|\alpha|\e^{\rmi\theta}$.

For the angular part of the basis we use the eigenfunctions of the
effective hole spin $J=I+S_{\rm h}$, the effective angular momentum
$F=L+J$, and the electron spin $S_{\rm e}$.
At the $\Gamma$ point, $J$ is a good quantum number and distinguishes
between the yellow exciton series ($J = 1/2$) and the green exciton
series ($J=3/2$).
$F$ and $S_{\rm e}$ are coupled to $F_t=F+S_{\rm e}$ with $z$
component $M_{F_t}$.
The complete basis set is then given by \cite{Schweiner16b,Schweiner17a}
\begin{equation}
  \ket{\Pi} = \ket{N,L;(I,S_h),J;F,S_e;F_t,M_F} \, .
\label{eq:basis}
\end{equation}
To obtain a finite size basis for the numerical computations the
quantum numbers must be restricted.
For each value of the principal quantum number $n=N+L+1$ we use
\cite{Schweiner16b}
\begin{eqnarray}
  L &= 0,\dots,n-1,\nonumber\\ 
  J &= 1/2, 3/2,\nonumber\\ 
  F &= |L-J|,\dots,\min(L+J,F_{\max}),\nonumber\\ 
  F_t &= F-1/2, F+1/2,\nonumber\\ 
  M_{F_t} &= -F_t,\dots,F_t \, .
\label{eq:quantum_numbers}
\end{eqnarray}
The excitonic wave functions can be expanded in the
basis \eref{eq:basis} as
\begin{equation}
  \ket{\Psi} = \sum_{NLJFF_tM_{F_t}}c_{NLJFF_tM_{F_t}}\ket{\Pi} \, , 
\label{eq:Psi}
\end{equation}
with the coefficients $\bi{c}$.
Using the Hamiltonian \eref{eq:Hamiltonianmitfelder} and the basis set
\eref{eq:basis} we can now set up the generalised eigenvalue problem
\begin{equation}
  \bi{Dc} = E\bi{Mc}
\label{eq:eigenwertproblem}
\end{equation}
for the resonance energies $E$ and the coefficients $\bi{c}$ of the
corresponding wave functions \eref{eq:Psi}.
The matrix elements of the matrices $\bi{D}$ and
$\bi{M}$ are given in the appendices of references
\cite{Schweiner16b,Schweiner17a} with the only difference that
$\alpha$ is now a complex parameter $\alpha=|\alpha|\e^{\rmi\theta}$
with $\theta$ the angle of the complex coordinate-rotation, as
explained above.
The matrix $\bi{M}$ in \eref{eq:eigenwertproblem} is the overlap
matrix of the basis states \eref{eq:basis} and differs from the
identity matrix because, as mentioned, the Coulomb-Sturmian functions
\eref{eq:basis_r} are not orthogonal.

Note that both matrices $\bi{D}$ and $\bi{M}$ in
\eref{eq:eigenwertproblem} are complex symmetric but non-Hermitian
matrices.
The generalised eigenvalue problem \eref{eq:eigenwertproblem} can be
solved numerically by application of the QZ algorithm, which is
implemented in the LAPACK routine ZGGEV \cite{LAPACK}.
To achieve convergence of the eigenvalues and eigenvectors the
maximum value for $n$ and the value $F_{\max}$ for the setup
\eref{eq:quantum_numbers} of the basis must be chosen sufficiently
large.
The LAPACK routine does not provide normalised eigenvectors.
For the computation of oscillator strengths in the next section
\ref{sec:Oscillatorstrength} the wave functions \eref{eq:Psi} must be
normalised according to
\begin{equation}
  \braket{\Psi_i|\bi{M}|\Psi_j} = \delta_{ij} \, ,
\label{eq:norm}
\end{equation}
which is achieved with a modified Gram-Schmidt process.
 
\subsection{Oscillator strengths}
\label{sec:Oscillatorstrength}
With the eigenvalues and eigenvectors obtained by numerical
diagonalisation of the generalised eigenvalue
problem~\eref{eq:eigenwertproblem} we are able to calculate 
the oscillator strengths for dipole transitions.
% added note, please check
Note that the crystal ground state depends on Bloch functions, which
are not explicitly known, and therefore only relative oscillator
strengths can be computed~\cite{Schweiner17c}.
% end added note
For circularly polarised light the relative
oscillator strength is given by \cite{Schweiner17a}
\begin{equation}
  f_{\rm{rel}}\sim\left(\lim_{r \rightarrow 0}
    {\frac{\partial}{\partial r}\braket{\sigma^\pm_z|\Psi(\bi{r})}}\right)^2 
\label{eq:frel}
\end{equation} 
with 
\begin{eqnarray}
\label{eq:sigma_z^+}
\ket{\sigma^+_z}=\frac{-i}{\sqrt{2}}\left(\ket{\pi_x}+i\ket{\pi_y}\right)=\ket{2,-1}_D,\\
\label{eq:sigma_z^-}
\ket{\sigma^-_z}=\frac{i}{\sqrt{2}}\left(\ket{\pi_x}-i\ket{\pi_y}\right)=-\ket{2,1}_D,
\end{eqnarray}
and
\begin{eqnarray}
%\label{xrichtung}
\ket{\pi_x}=\frac{i}{\sqrt{2}}\left(\ket{2,-1}_D+\ket{2,1}_D\right),\\
%\label{yrichtung}
\ket{\pi_y}=\frac{1}{\sqrt{2}}\left(\ket{2,-1}_D+\ket{2,1}_D\right),
\end{eqnarray}
for an electric and/or magnetic field in $[001]$ direction.
We use the abbreviation $\ket{F_t,M_{F_t}}_D$ to denote the states
\begin{eqnarray}
\ket{F_t,M_{F_t}}_D &= \ket{(S_{\rm e},S_{\rm h})\,S,\,I;\,I+S,\,L;\,F_t,\,M_{F_t}}\nonumber\\
&=\ket{(1/2,1/2)\,0,\,1;\,1,\,1;\,F_t,\,M_{F_t}} \, ,
\end{eqnarray}
where the coupling scheme differs from the one given in section \ref{sec:eigenvalues}.
The spins couple in the following way \cite{Schweiner17a}:
\begin{equation}
  S_{\rm e}+S_{\rm h} = S \rightarrow (I+S)+L=F_t\,,
\end{equation}
with the total spin $S$, the quasispin $I$, the angular momentum $L$,
the total angular momentum $F_t$ and its projection on the
quantisation axis $M_{F_t}$.
With the relative oscillator strength we can furthermore calculate the
spectrum.
Rescigno and McKoy \cite{McKoy} have shown that the photoabsorption
cross section in atomic physics, using the complex
coordinate-rotation, can be written as
\begin{equation}
  \sigma(E) = 4\pi\alpha(E-E_0)\,\mathrm{Im}
  \sum_j\frac{\braket{\Psi_0|D|\Psi_j(\theta)}^2}{E_j-E},
\label{eq:sigma}
\end{equation} 
where $\alpha$ is the fine-structure constant, $E_0$ the energy of
the ground state $\Psi_0$, and $E_j$ the complex energy of the
resonance state $\Psi_j$.
For excitonic spectra we replace the squared dipole matrix elements
$\braket{\Psi_0|D|\Psi_j(\theta)}^2$ in \eref{eq:sigma} with the
relative oscillator strengths $f_{\mathrm{rel}}$ given in \eref{eq:frel}.
The excitonic absorption spectrum then reads
\begin{equation}
  f(E) = -\frac{1}{\pi}{\mathrm{Im}}\sum_j\frac{f_{\rm rel}^{(j)}}{E-E_j}
\label{eq:Spektrum}
\end{equation}
with $E_j$ the complex energies of the resonances.   
Note that the squared dipole matrix elements
$\braket{\Psi_0|D|\Psi_j(\theta)}^2$ in \eref{eq:sigma} and the
relative oscillator strength $f_{\mathrm{rel}}$ in \eref{eq:frel} are
real-valued for bound states but can become complex for resonances,
because, after complex coordinate-rotation, bra vectors are not the
complex conjugate of the ket vectors (see~\cite{Rei82,Ho83,Moi98}),
as in Hermitian quantum mechanics.
The complex phases of $f_{\mathrm{rel}}$ lead to deviations of
resonance shapes from a Lorentzian profile in \eref{eq:Spektrum}.

\section{Results and discussion}
\label{sec:results}
In this section we present the results of our calculations for
excitons of cuprous oxide in external fields.
We investigate the resonances in the complex energy plane and the
corresponding absorption spectra obtained with circularly polarised
light for excitons, first in electric fields, and then in parallel
electric and magnetic fields.
We restrict the presentation of results to resonances and ignore the
bound states, which would appear as delta peaks in the absorption
spectra.
The linewidths of resonances in our calculations are solely caused by
the external electric field, i.e., we do not consider exciton-phonon
interactions \cite{Schweiner16a,Stolz18}.
\begin{figure}
\includegraphics[width=0.49\textwidth]{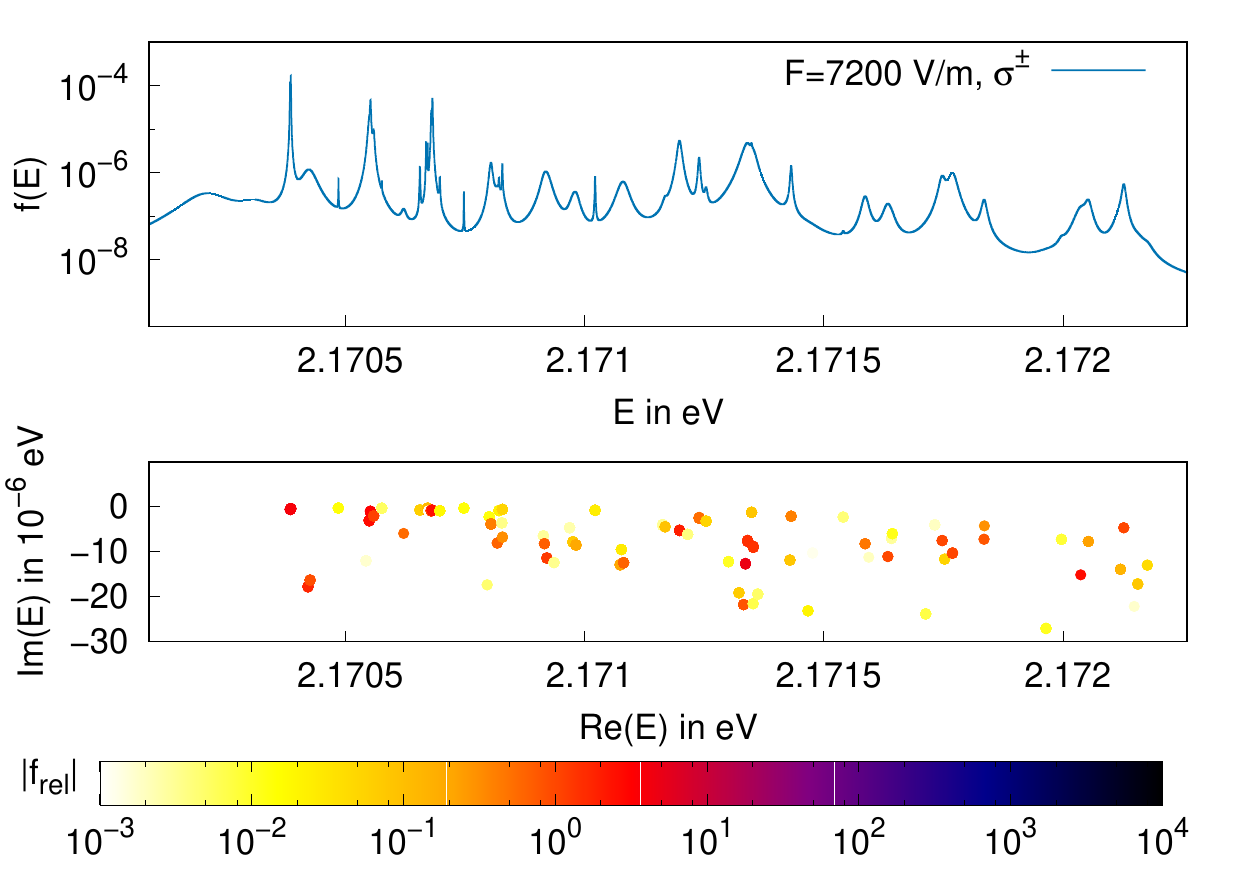} %{7200.pdf}
\caption{Lower part: Positions of resonances in the complex energy
  plane for excitons in cuprous oxide in an electric field with field
  strength $F=7200\,$V/m in [001] direction.  The colours of the
  symbols encode the absolute values of the relative oscillator
  strengths $f_{\mathrm{rel}}$ for excitations with $\sigma^\pm$
  polarised light.  Upper part: The corresponding absorption spectra
  for excitations with $\sigma^+$ and $\sigma^-$ polarised light coincide.}
\label{fig2:position_F=7200}
\end{figure}
\begin{figure}
\includegraphics[width=0.49\textwidth]{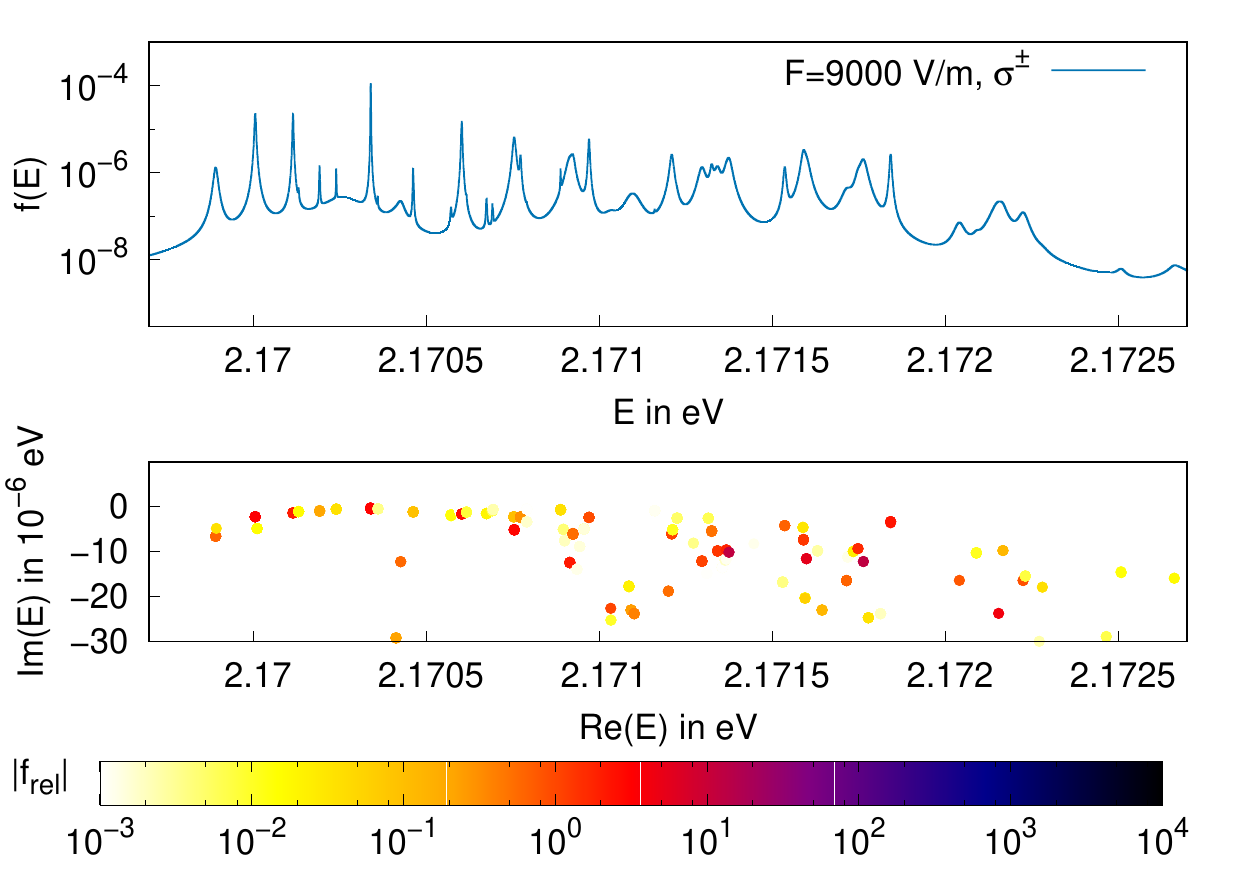} %{9000.pdf}
\caption{Same as figure~\ref{fig2:position_F=7200} but for $F=9000\,$V/m.}
\label{fig3:position_F=9000}
\end{figure}

\subsection{Electric fields in [001] direction}
In figures~\ref{fig2:position_F=7200} and \ref{fig3:position_F=9000}
we present the results for excitons in an electric field oriented
along the [001] axis with field strengths $F=7200\,$V/m and
$F=9000\,$V/m, respectively.
The lower parts of the figures show the positions of resonances in the
complex energy plane obtained as complex eigenvalues of the
non-Hermitian generalised eigenvalue problem~\eref{eq:eigenwertproblem}.
For the computations we used the basis~\eref{eq:basis} with principal
quantum numbers up to $n=30$ and $F_{\max}=15$ resulting in a total
set of $5303$ basis functions.
For the complex coordinate-rotation we used rotation angles in the
region $0.1 < \theta < 0.3$.
The colours of the resonance positions encode the absolute values of
the relative oscillator strengths $f_{\mathrm{rel}}$ for excitations
with circularly polarised light given by \eref{eq:frel}.
The upper parts of figures~\ref{fig2:position_F=7200} and
\ref{fig3:position_F=9000} show the absorption spectra $f(E)$ obtained
using \eref{eq:Spektrum}.
Note that the absorption spectra for $\sigma^+$ and $\sigma^-$
polarised light coincide as expected.

The spectrum at $F=7200\,$V/m, in figure~\ref{fig2:position_F=7200},
exhibits resonances with quite different linewidths.
Long-lived resonances appear as thin peaks in the absorption spectra.
In general, broader peaks belong to resonances with higher principal
quantum numbers $n$ or, within a given $n$-manifold, to resonances
with lower energy \cite{Heck18}.
The resonances shown in the figure belong to principal quantum
numbers between $n=8$ and $n=15$.
Note that, for the chosen electric field strength, the different
$n$-manifolds already strongly overlap.
If we increase the electric field strength to $F=9000\,$V/m, new
long-lived resonance states appear (see figure~\ref{fig3:position_F=9000})
and the lifetimes of the resonances with higher real energy part
decrease.
If we compare the positions in both plots, we recognise that
more resonances appear deeper in the lower half of the complex energy
plane.
Undoubtedly, the reason for this is the electric field which
lowers the potential barrier of the Coulomb potential and therefore
increases the tunnel probability.

\begin{figure}
\includegraphics[width=0.5\textwidth]{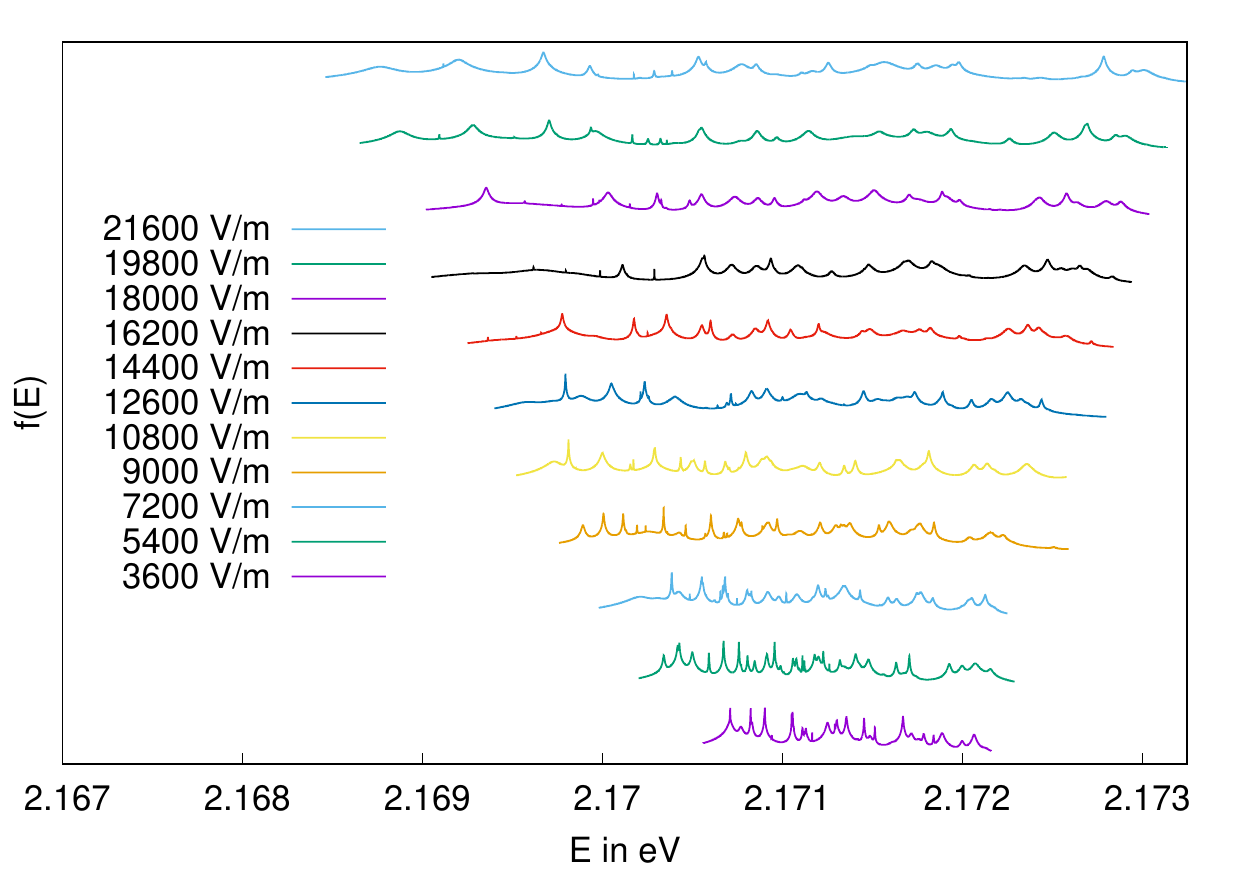} %{waveplot.pdf}
\caption{Evolution of resonance spectra for $\sigma^\pm$ polarised
  light as a function of the electric field.  The electric field
  strength increases from $F=3600\,$V/m (bottom) to $F=21600\,$V/m
  (top), and is orientated in the [001] direction.}
\label{fig4:waveplot}
\end{figure}

Figure~\ref{fig4:waveplot} shows the absorption spectra for different
electric field strengths from $F=3400\,$V/m to $F=21600\,$V/m.
The single spectra $f(E)$ are plotted with an offset, therefore no
units of $f$ are given.
For the computations we have used the same basis set as described
above.
The figure shows the evolution of the spectra in dependence on the
electric field strength.
All spectra are restricted to the range where resonances appear.
At lower energies only bound states would appear and at higher ones
the spectra would show unconverged states due to the finite basis.
We first notice the fan-like spreading of the absorption spectra.
That means for $F=3600\,$V/m we found resonances between $2.17056\,$eV
and $2.17216\,$eV and for $F=21600\,$V/m we found resonances between
$2.16846\,$eV and $2.17324\,$eV.
This behaviour derives from the Stark effect which splits the energy
levels and moves the positions of the resonances along the real axis.
Additionally, the decrease of the potential barrier moves the
resonances deeper into the lower half of the complex plane.
Therefore we observe mostly resonances with short lifetime (broad peaks)
for field strengths $F>14400\,$V/m.
New lines always appear as sharp peaks (long-lived resonances) on the
left hand side of the absorption spectra.
For higher field strengths the number of exposed resonances decreases
because the maximum value of $\theta$ is too small.
In principle, we could increase $\theta$ to uncover these states,
however, in that case the basis set must be increased, which leads to
higher computation times.

In reference~\cite{Heck18} it has been shown that the field strength
for the dissociation of excitons in Cu$_2$O decreases with increasing
principal quantum number $n$, but increases, for fixed $n$, with
growing exciton energy, in agreement with similar results for the
Stark effect in atoms.
We expect a similar behaviour in our spectra, however, this can not
easily be observed because we are in an energy and field strength
region, where states with different $n$ strongly overlap.
In particular, we have not yet been able to assign any (approximate)
quantum numbers to the resonances shown in figure~\ref{fig4:waveplot}.
To do so, e.g.\ the evolution of resonance spectra in
figure~\ref{fig4:waveplot} must be followed on a much denser grid of
field strengths down to the field-free spectrum.
As the computation of each spectrum is numerically very expensive this
is currently beyond our numerical capabilities.

\subsection{Parallel electric and magnetic fields}
Now we investigate the influence of an additional magnetic field
parallel to the electric field.
Both fields are in [001] direction.
\begin{figure}
\includegraphics[width=0.49\textwidth]{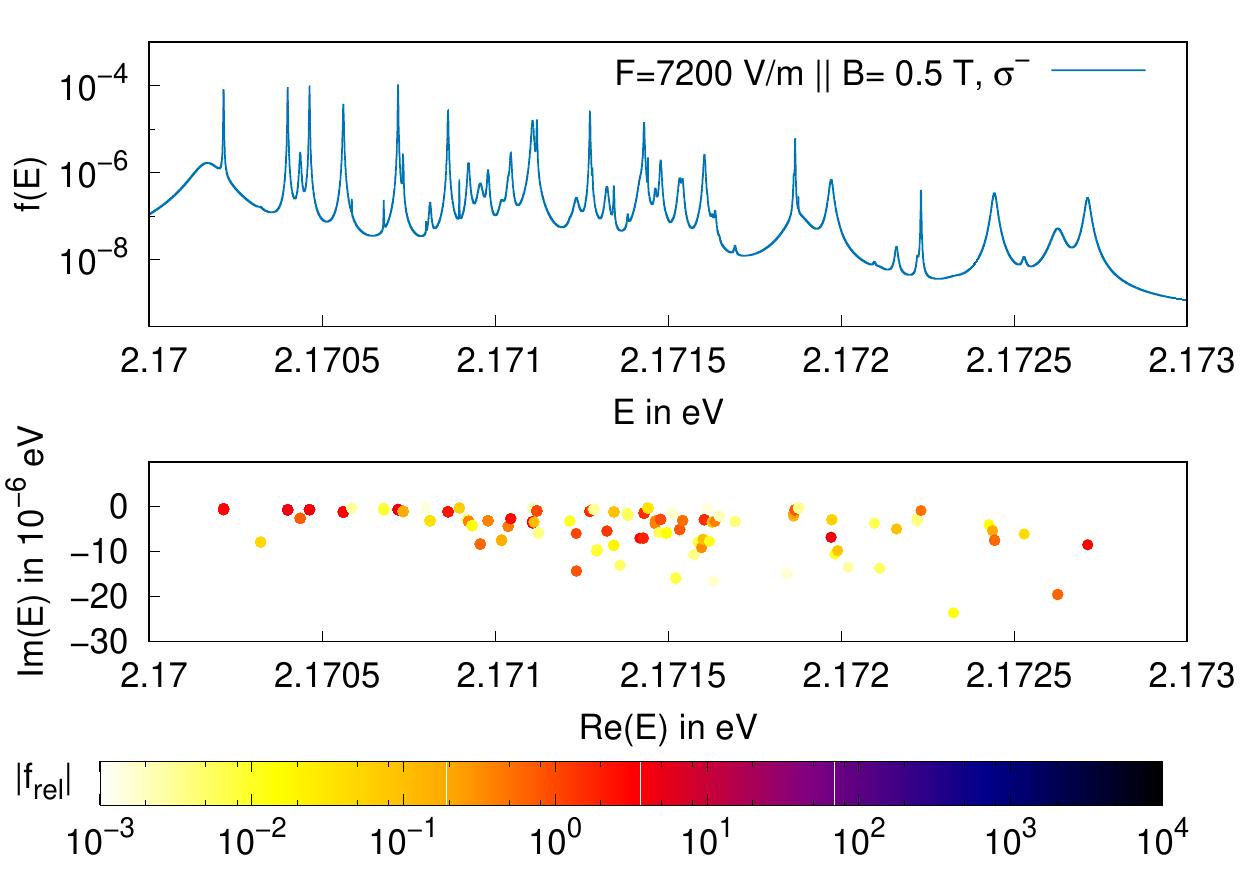} %{7200_sminus_05T.pdf}
\caption{Position and resonance spectra for an electric field
  $F=7200\,$V/m and magnetic field $B=0.5$ T. Both fields are in [001]
  direction and the light is $\sigma^-$ polarised.}
\label{fig5:position_F=7200_05T_sminus}
\end{figure}
\begin{figure}
\includegraphics[width=0.49\textwidth]{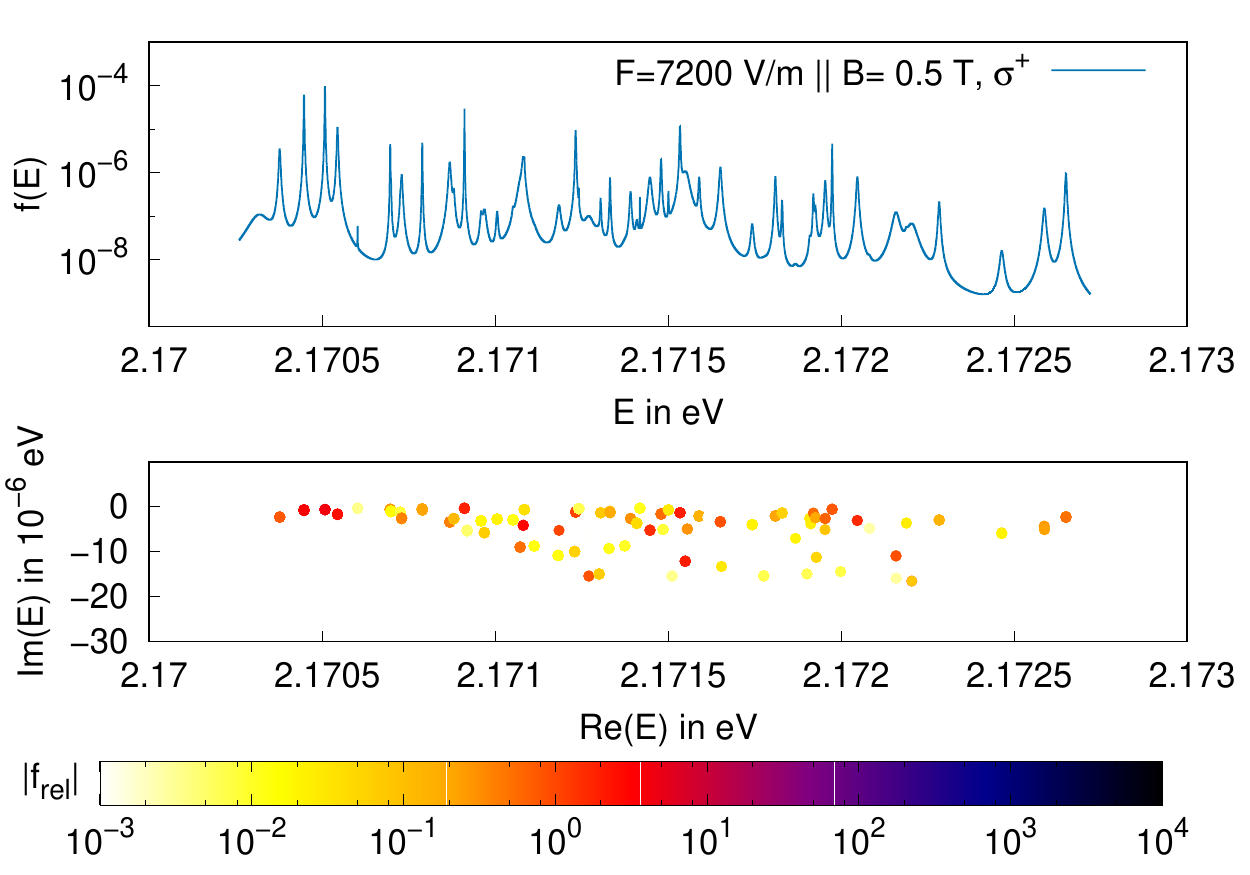} %{7200_splus_05T.pdf}
\caption{Same as figure~\ref{fig5:position_F=7200_05T_sminus} but for
  $\sigma^+$ polarised light.}
\label{fig6:position_F=7200_05T_splus}
\end{figure}
Figure~\ref{fig5:position_F=7200_05T_sminus} shows the
positions of the resonances in the complex energy plane and the
absorption spectrum for $F=7200\,$V/m, $B=0.5\,$T and $\sigma^-$
polarised light.
For the same field strength but with $\sigma^+$ polarised light,
figure~\ref{fig6:position_F=7200_05T_splus} shows the absorption
spectrum and the positions of resonances in the complex energy plane.
As discussed above, both spectra consist of long-lived resonances,
thin peaks, and short-lived ones, broad peaks, but if we compare the
spectra of $\sigma^-$ and $\sigma^+$ polarised light (see
figure~\ref{fig6:position_F=7200_05T_splus}) they differ from each
other.
This is due to the magnetic field and the symmetry of the valence band.
In \cite{Schweiner17a} it is shown that different excitons are excited
by $\sigma^-$ and $\sigma^+$ polarised light with a magnetic field in
[001] direction.
With $\sigma^+$ polarised light excitons with a large amount of
angular momentum $L=1$, $F_t=2$, $M_{F_t}=-1$ and with $\sigma^-$
polarised light excitons with a large amount of angular momentum
$L=1$, $F_t=2$, $M_{F_t}=+1$ (cf.\ equations (\ref{eq:sigma_z^+}) and
(\ref{eq:sigma_z^-})) are strongly excited.
These states are non-degenerate in a magnetic field or in parallel
magnetic and electric fields.
We note that, as for the Stark spectra discussed above, we are not
able to assign any quantum numbers to individual resonances.

\section{Conclusion and outlook}
\label{sec:conclusion}
Schweiner \etal\cite{Schweiner16b,Schweiner17a} have developed a
method for the numerically exact computation of yellow excitons in
cuprous oxide by using a complete basis set.
We have extended and augmented this technique by application of the
complex-coordinate-rotation method, which, as a novel result, allows
for the computation of unbound excitonic resonance states.
We have used the method to calculate the positions of resonances in
the complex energy plane, and thus the decay rates, for excitons in
external electric fields and in parallel electric and magnetic fields.
Furthermore, we have simulated the absorption spectra for excitations
with circularly polarised light, and have shown that the spectra
obtained with $\sigma^+$ and $\sigma^-$ polarised light coincide for
excitons in an electric field but significantly differ for excitons in
combined electric and magnetic fields.

The computation of magnetoexcitons including the effects of the
valence band have allowed for detailed line-by-line comparisons
between experimental and theoretical spectra \cite{Schweiner17a}.
A detailed comparison of our theoretical excitonic resonances with
experimental absorption spectra \cite{Heck18} will be an interesting
future task and will clarify the validity or limitations of the
hydrogenlike model, with its refinements.

An interesting property of the complex generalised eigenvalue
problem~\eref{eq:eigenwertproblem} is that for certain values of the
electric and magnetic field strengths the resonance energies and also
the corresponding eigenstates can become degenerate.
This situation is not possible in Hermitian quantum mechanics, and is
called an \emph{exceptional point} \cite{Kato66,Hei90,Moi11}.
Such points have been found in computations for the hydrogen atom in
combined electric and magnetic fields, however, at very high and thus
experimentally not accessible field strengths \cite{Car07,Car09,Fel16}.
With the method introduced in this paper it will be possible to search
for exceptional points in the spectra of cuprous oxide and in regions
of the field strengths, which can easily be realised in experiments.
Cuprous oxide could therefore be an excellent candidate for the first
experimental observation of an exceptional point in a Rydberg system.

Nikitine \cite{Nikitine1959} has investigated experimentally the green
exciton series in Cu$_2$O and, recently, Kr\"uger and Scheel
\cite{KruegerInterseries2019} have focused on the interseries
transitions, e.g., between yellow and green excitons.
In this context, a better understanding of the green exciton series is
desirable.
Since the green series is located inside of the yellow continuum
\cite{Nikitine1961,Nikitine1963,MALERBA20112848}, and the different series couple, the green
exciton states are actually resonances.
The complex coordinate-rotation method used in this paper thus is also
an appropriate tool for the future investigation of these resonance
states.

\ack
This work was supported by Deutsche Forschungsgemeinschaft (DFG)
through Grant No.~MA1639/13-1.
We thank G.~Wunner for a careful reading of the manuscript.

\section*{References}

\bibliographystyle{unsrt}
%\bibliography{paper}

\end{document}